\documentclass[final,english]{aipproc}

 \layoutstyle{8x11double}
\usepackage[english]{babel}

\begin{document}

\title{How Efficient is Rotational Mixing in Massive Stars ?}




\author{I. Brott}{
  address={Sterrenkundig Instituut Utrecht, Princetonplein 5,
    3584 CC Utrecht, The Netherlands}
}

\author{I. Hunter}{
  address={Department of Physics and Astronomy, The Queen's University of
    Belfast, BT7 1NN,  Northern Ireland, UK}
}

\author{P. Anders}{
  address={Sterrenkundig Instituut Utrecht, Princetonplein 5,
    3584 CC Utrecht, The Netherlands}
}

\author{N. Langer}{
  address={Sterrenkundig Instituut Utrecht, Princetonplein 5,
    3584 CC Utrecht, The Netherlands}
}

\begin{abstract}
    The VLT-Flames Survey for Massive Stars \cite[]{Evans05,Evans06} provides precise measurements of
    rotational velocities and nitrogen surface abundances of massive 
    stars in the Magellanic Clouds. Specifically, for the first time, such abundances have been estimated
    for stars with significant rotational velocities. This extraordinary data set
    gives us the unique possibility to calibrate rotationally and magnetically induced
    mixing processes. Therefore, we have computed a grid of stellar evolution
    models varying in mass, initial rotational velocity and chemical composition. In our 
    models we find that although magnetic fields generated by the
    Spruit-Taylor dynamo are essential to understand the
    internal angular momentum transport (and hence the rotational behavior),
    the corresponding chemical mixing must be neglected
    to reproduce the observations. Further we show that for low metallicities 
    detailed initial abundances are of prime importance, as solar-scaled
    abundances may result in significant calibration errors.
\end{abstract}

\maketitle


\section{Introduction}
Rotational velocities and nitrogen abundances have been measured
for a large sample of LMC stars as part of the VLT-FLAMES Survey
of Massive Stars  \cite[]{Evans05,Evans06}. These data have been used to set constraints on
the efficiency of rotational induced 
mixing processes in our models. With these newly calibrated models we have
investigated the influence of different chemical mixtures at SMC metallicity on
the models.

\section{The Models}
We have calculated models using a stellar evolution code for single and binary stars 
\cite[]{Heger00,Langer98}. The models include rotation and magnetic
fields. We have extended the code to account for different initial chemical
compositions. For mass loss we have adopted the prescription of
\cite{Vink00,Vink01}.\\
\begin{table}[!b]
\begin{tabular}{lcccccc}
\hline\hline
    \tablehead{1}{r}{b}{Mixture}
  & \tablehead{1}{r}{b}{C}
  & \tablehead{1}{r}{b}{N}
  & \tablehead{1}{r}{b}{O}
  & \tablehead{1}{r}{b}{Mg}
  & \tablehead{1}{r}{b}{Si}
  & \tablehead{1}{r}{b}{Fe}\\
\hline
Grev96 & 8.55 & 7.97 & 8.87 & 7.58 & 7.55 & 7.50 \\
Aspl05 & 8.39 & 7.78 & 8.66 & 7.53 & 7.51 & 7.45 \\
LMC & 7.75 & 6.90 & 8.35 & 7.05 & 7.20 & 7.05 \\
SMC & 7.37 & 6.50 &7.98 & 6.72 & 6.80 & 6.78 \\
\hline\hline
\end{tabular}
\caption{\bf Baseline abundances of C, N, O, Mg, Si and Fe used for the
  different chemical mixtures.}
\label{tab:mixtures}
\end{table}
We have compiled chemical mixtures for the Small and Large Magellanic Cloud  
and compared them to solar scaled mixtures based on  Grevesse96 \cite{Grevesse96} and
Asplund05 \cite{Asplund05}.
The LMC and SMC mixtures are based on the Asplund composition in which the
amount of all metals has been decreased by $0.4\,\rm{dex}$ and
$0.7\,\rm{dex}$, respectively, and the abundances of C, N, O, Mg, Si and Fe have
been set to the baseline values (C. Trundle ,private communications) summarized 
in Table \ref{tab:mixtures}.

\begin{figure}
  \includegraphics[height=.34\textheight,angle=-90]{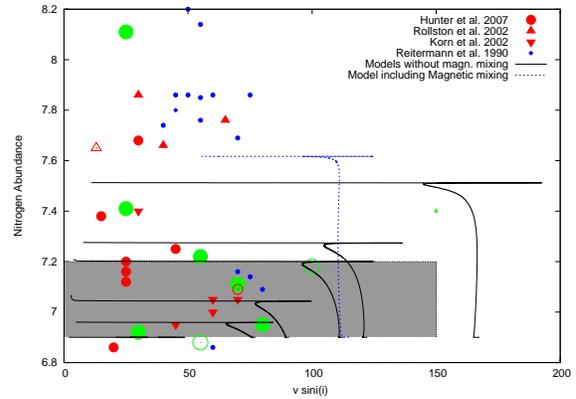}
\caption{Projected surface velocity versus nitrogen abundance of LMC B-Stars. Red (middle sized),
  green (large sized), blue (small sized)  symbols represent dwarfs ($\log g < 3.7 $), giants ($3.7 \leq
  \log g \leq 3.2$), supergiants ($\log g < 3.2$), respectively. Open symbols represent upper limits.
  The gray   shaded box represents the bulk of stars in \cite{Hunter07a}. Black solid lines represent
 models of $13 \rm{M}_\odot$ at different initial velocities which are calibrated to fit the data of \cite{Hunter07a}. 
The model which is plotted with the blue dotted line takes chemical mixing processes due to magnetic fields into account.}
\label{fig:vsin-lmc}
\end{figure}

\begin{figure}
  \begin{tabular}{c}
  \includegraphics[height=.32\textheight,angle=-90]{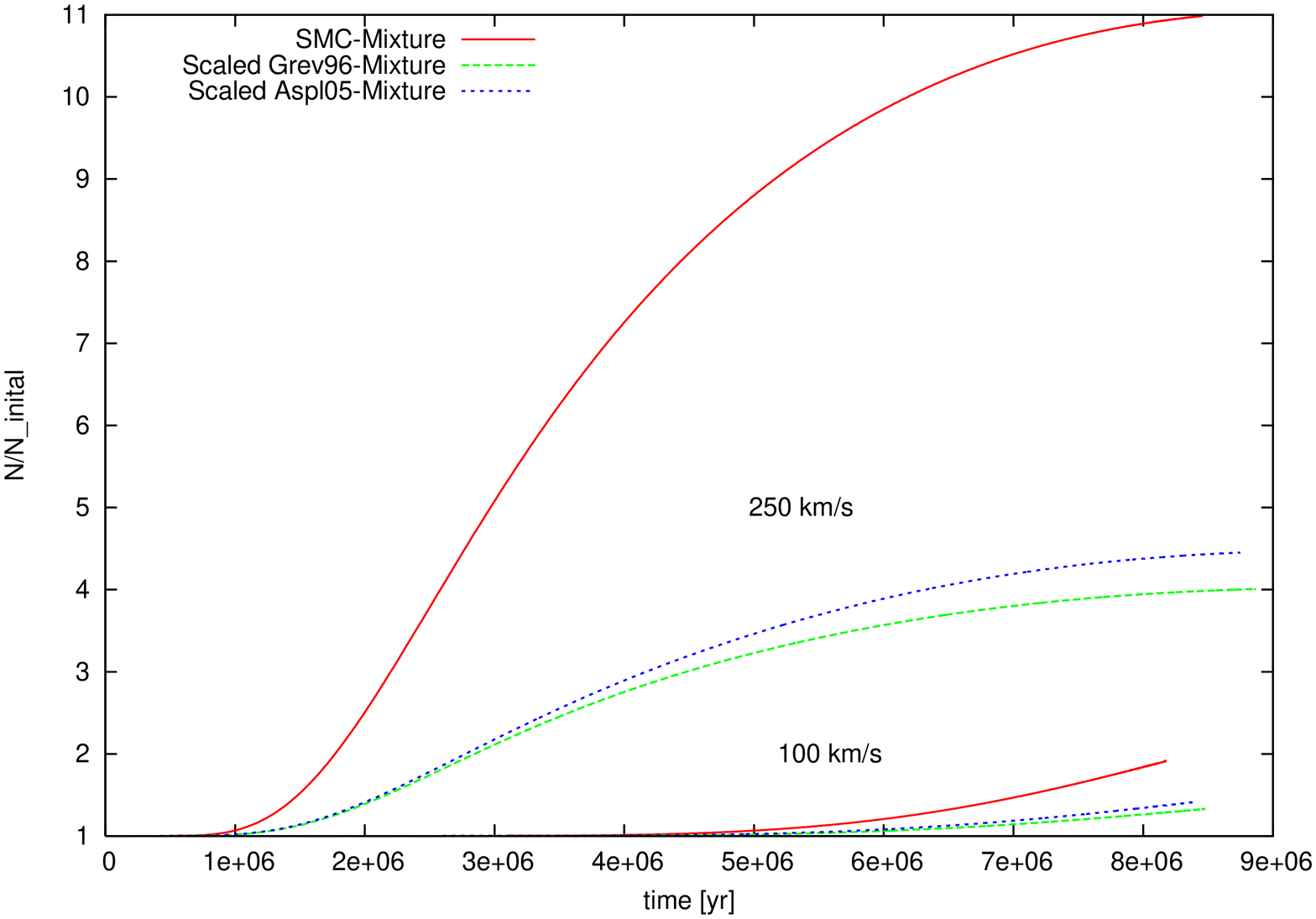}\\
  \includegraphics[height=.32\textheight,angle=-90]{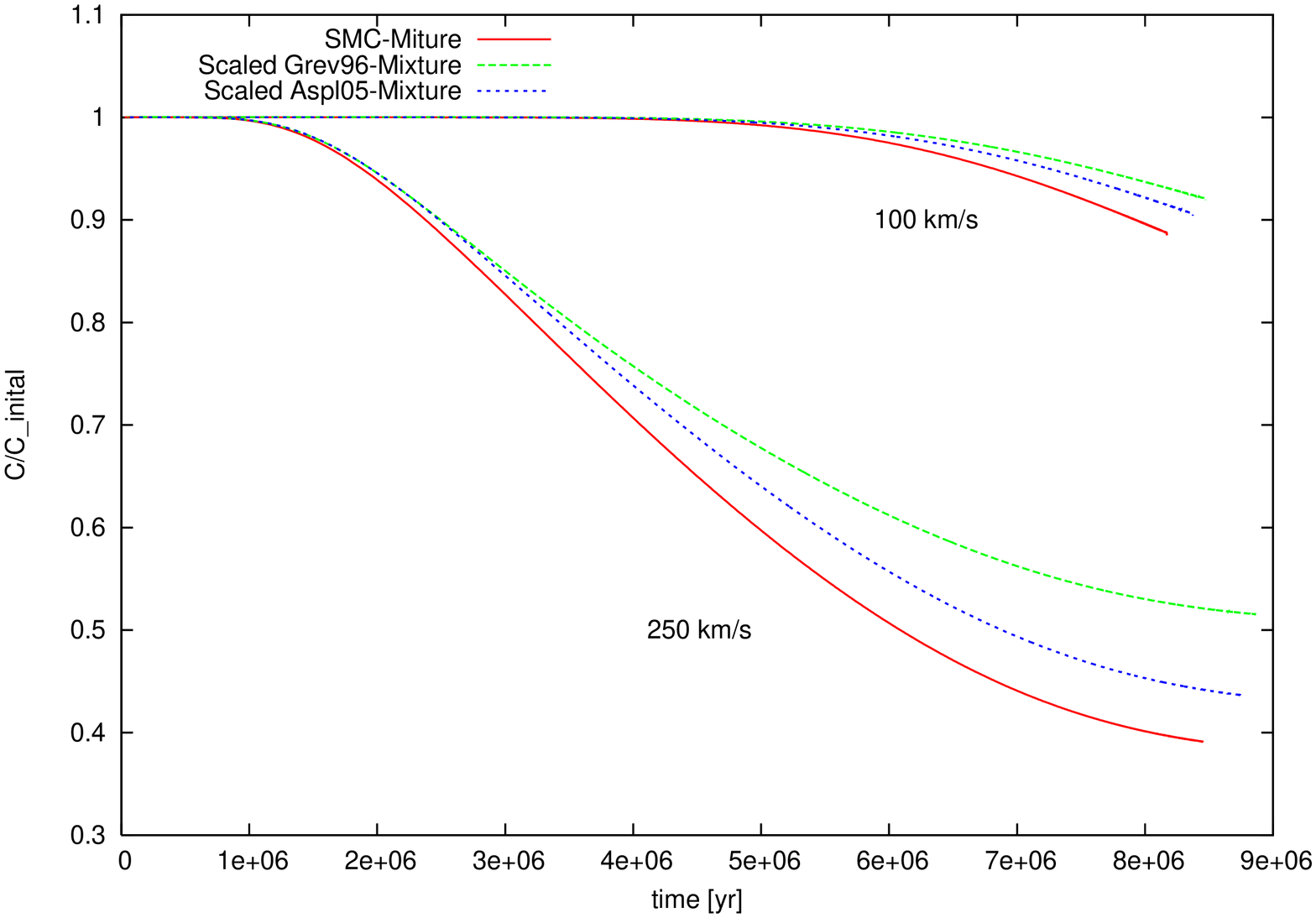}\\
  \includegraphics[height=.32\textheight,angle=-90]{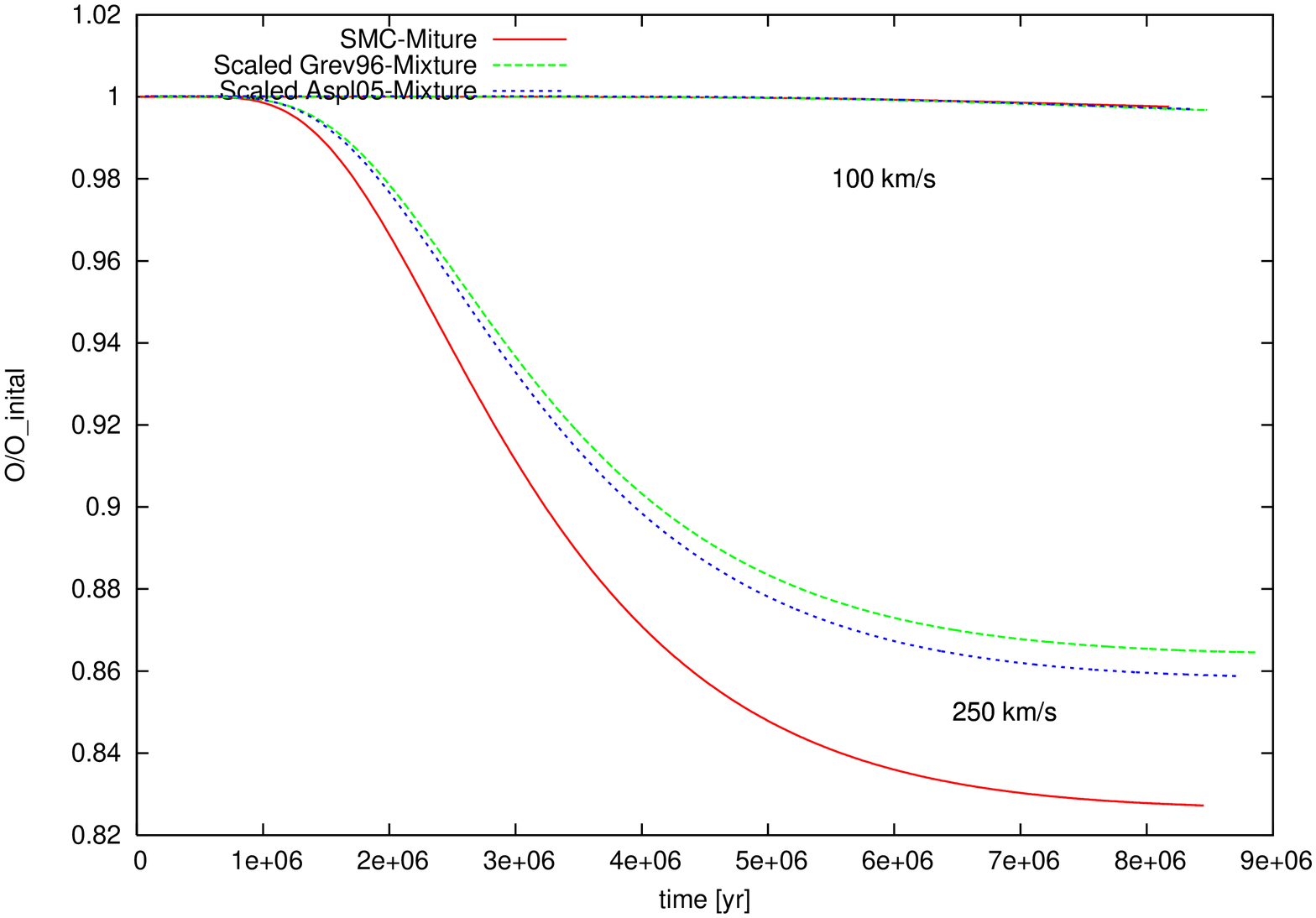}\\
  \end{tabular} 
  \caption{CNO surface abundance as a function of time for models with SMC mixture (red solid line) and 
    solar-scaled SMC-Models based on the Grevesse96 mixture (green dashed line) and Asplund05 
    mixture (blue dotted).
     From top to bottom, enrichment in N, depletion in C and O for
    $20 \rm{M}_\odot$ models with 100 and 250\,km/s initial velocity. Higher
    velocity results in higher enrichment/depletion.} 
  \label{fig:cno-enrichment}
\end{figure}

We have used the nitrogen abundances and rotational velocities of the LMC,
measured by the Flames-Survey \cite[in preparation]{Hunter07a}, to
calibrate the chemical mixing efficiency $f_c$ for rotationally induced
mixing processes \cite[]{Heger00}. 
We have used a model of $13\rm{M}_\odot$  and  an initial rotational velocity of
140 km/s (at the Zero Age Main Sequence) which agrees with the average of the sample.
The chemical mixing efficiency $f_c$ has been calibrated such that the nitrogen surface abundance reached
7.2 dex at the end of the main sequence lifetime, which gave $f_c=2.28\cdot10^{-2}$. 
The models assume an overshooting parameter of $\alpha=0.335$, which has been
adjusted to fit the observations \cite{Hunter07b}. \\
Magnetic fields are treated as described in \cite{Spruit02,Heger05}. 
Even though magnetic fields are needed to transport angular momentum, the
efficiency for the chemical mixing must be very small. 
In Fig. \ref{fig:vsin-lmc}, the black solid lines represent models where chemical mixing due to magnetic fields is disabled.  
For comparison the model which is plotted with the blue dotted line shows the nitrogen abundances reached if
chemical mixing due to magnetic fields is included. The chemical diffusion process is then dominated
by magnetic diffusion, resulting in too strong mixing, incompatible with the observations. The data points
show data of \cite{Hunter07} (circles), \cite{Rolleston02} (triangles),
\cite{Korn02} (upside-down triangles) and \cite{Reitermann90} (diamonds). The
grey box indicates data that will become available soon \cite[in preparation]{Hunter07a}.  

\section{Solar-Scaled Models gone bad ? }
\begin{figure}[b]
 \includegraphics[height=.34\textheight,angle=-90]{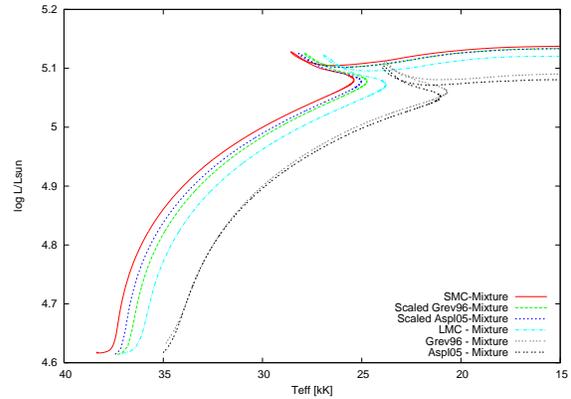}
 \caption{Evolutionary tracks of $20\rm{M}_\odot$ models at 100\,km/s initial
  velocity. The colors/line codings describe different mixtures for the initial composition used.}
\label{fig:smchrd}
\end{figure}
To reproduce observed abundances and to understand how rotational induced
mixing processes work in massive stars, it is of key importance to take the
correct initial chemical composition into account. \\
To investigate the effect of the initial chemical composition, we compare the
chemical enrichment/depletion of CNO in a
$20\,\rm{M}_\odot$ model at 100 and 250\,km/s initial velocity and three
different chemical mixtures. \\
Fig. \ref{fig:cno-enrichment} shows the chemical enrichment of N and the depletion in C and O
during the main-sequence evolution for a SMC mixture and  scaled-solar abundance patterns of Grevesse96
 and Asplund05 for comparison. The solar mixtures have been scaled such that the Fe
abundance is the same as for the SMC mixture. 
The CNO baseline abundances for the mixtures are
given in Table \ref{tab:smcbaseline}. In the case of nitrogen, the baseline abundance can differ
(depending on the model) by up to 0.7 dex. 
The enrichment in nitrogen for the SMC mixture is about $\sim$4 times higher than for
solar scaled models. For C and O the effect is less pronounced, but the depletion
in C is still up to a factor 2 larger in the SMC mixture than in the solar-scaled
models. We assume, the reason is the larger C/N-ratio of the SMC composition.
Therefore, on the onset of the CNO cycle the model contains more carbon which can be 
converted into nitrogen and mixed into the bottom of the envelope 
before the increasing H-He gradient avoids more nitrogen leaving the core.
\\

The HR-diagram in Fig.\ref{fig:smchrd} shows evolutionary tracks for a model of $20\rm{M}_\odot$ at
100\,km/s. Shown are models (from left to right) for SMC, scaled Grevesse96,
scaled Asplund05 mixtures and for comparison LMC and unscaled Grevesse96 and
Asplund05 mixtures. One can see the typical shift to higher temperatures for 
the models with decreasing metallicity. A track using the SMC mixture is about
1\,kK hotter than a comparable solar scaled model, but the temperature
difference between the LMC-mixture and a solar-scaled SMC mixture is of the
same order. 

\begin{table}
\begin{tabular}{lcccc}
\hline\hline
  \tablehead{1}{r}{b}{Mixture}
& \tablehead{1}{r}{b}{C}
& \tablehead{1}{r}{b}{N}
& \tablehead{1}{r}{b}{O}
& \tablehead{1}{r}{b}{Fe}\\
\hline
Scaled-Grev96 & 7.83 & 7.25 & 8.15 & 6.78\\
Scaled-Aspl05 & 7.72 & 7.11 & 7.99 & 6.78\\
SMC &           7.37 & 6.50 & 7.98 & 6.78\\
\hline\hline
\end{tabular}
\caption{ Baseline abundances of C, N, O in SMC and Solar-Scaled-SMC Mixtures.}
\label{tab:smcbaseline}
\end{table}

\section{Conclusions \& Outlook}
\begin{itemize}
\item We find that magnetically induced instabilities  give too large diffusion
 coefficients for chemical mixing and must, therefore, be reassessed. 
\item We have investigated the influence of the initial chemical composition on the
  models. To reproduce the correct nitrogen abundances at low metallicity and 
  to be able to understand the ongoing mixing processes, solar-scaled initial
  abundances are not sufficient.
\item Next we have to investigate the influence of the mixture on the
  opacities, as the opacity determines temperature and luminosity of the
  model. Currently we are using OPAL95-Tables \cite[]{Iglesias96}, based on a
  Grevesse93 mixture \cite{Grevesse93}.
\item Rotating stars are known to show the effect of gravity darkening,
  which means that the poles become brighter than the equator
  region. The mass loss in the pole region will be therefore enhanced due to
  the larger luminosity, while mass loss at the equator region will be
  determined by rotation. Currently our models neglect this effect.
\end{itemize}





\bibliographystyle{aipproc}   

\bibliography{literatur}

\IfFileExists{\jobname.bbl}{}
 {\typeout{}
  \typeout{******************************************}
  \typeout{** Please run "bibtex \jobname" to optain}
  \typeout{** the bibliography and then re-run LaTeX}
  \typeout{** twice to fix the references!}
  \typeout{******************************************}
  \typeout{}
 }

\end{document}